\documentclass[final,3p,times]{elsarticle}
\usepackage{amsmath}
\usepackage{amssymb}
\usepackage{lineno,hyperref}
\usepackage{natbib}

\modulolinenumbers[5]
\biboptions{sort&compress}
\journal{Annals of Physics}
\hypersetup{colorlinks=true,linkcolor=blue,citecolor=blue,filecolor=blue,urlcolor=blue,breaklinks=true}
\RequirePackage{color}

\begin{document}

\begin{frontmatter}

\title{Fermions in the background of mixed vector-scalar-pseudoscalar
square potentials}

\author[mymainaddress]{Luiz P. de Oliveira\corref{mycorrespondingauthor}}
\cortext[mycorrespondingauthor]{Corresponding author}
\ead{luizp@if.usp.br}

\author[mysecondaryaddress]{Luis B. Castro}
\ead{lrb.castro@ufma.br}

\address[mymainaddress]{Instituto de F\'{\i}sica, Universidade de S\~{a}o Paulo (USP),
\\05508-900, S\~{a}o Paulo, SP, Brazil}
\address[mysecondaryaddress]{Departamento de F\'{\i}sica, Universidade Federal do Maranh\~{a}o, Campus Universit\'{a}rio do Bacanga,\\ 65080-805, S\~{a}o Lu\'{\i}s, MA, Brazil.}

\begin{abstract}
The general Dirac equation in 1+1 dimensions with a potential with a completely general Lorentz structure is studied. Considering mixed vector-scalar-pseudoscalar square potentials, the states of relativistic fermions  are investigated. This relativistic problem can be mapped into a effective Schr\"{o}%
dinger equation for a square potential with repulsive and attractive delta-functions situated at the borders. An oscillatory transmission coefficient is found and resonant state energies are obtained. In a special
case, the same bound energy spectrum for spinless particles is obtained, confirming the predictions of literature. We showed that existence of bound-state solutions are conditioned by the intensity of the pseudoscalar potential, which posses a critical value.
\end{abstract}


\end{frontmatter}


\section{Introduction}

Since its formulation in 1928 \cite{PRSLA117:610:1928}, the Dirac equation
has been widely investigated in various physical systems. Among the
applications, we can highlight the Klein's paradox \cite%
{ZP53:157:1929,JPA26:1001:1993,IJMPA14:631:1999}, the description of
electrons in graphene \cite{RMP81:109:2009}, the influence of the nuclear
medium on the nucleons \cite{WALECKA1986} and the relativistic hydrogen atom 
\cite{GREINER1990}.

The pseudospin symmetry (PSS) is a topic of intense discussion,
activities and recent progress in the last decades. PSS was
introduced in nuclear physics for explain the degeneracies of orbitals in
single particle spectra \cite{PLB30:517:1969,NPA137:129:1969}. In order to
understand the origin of PSS, we need to take into account the
motion of the nucleons in a relativistic mean field and thus consider the
Dirac equation \cite{PRL78:436:1997,PR315:231:1999,PR414:165:2005}. The case
in which the mean field is composed by a vector ($V_{t}$) and a scalar ($%
V_{s}$) potential, with $\Sigma =V_{t}+V_{s}=0$ ($V_{t}=-V_{s}$) is usually
pointed out as a necessary condition for occurrence of PSS in
nuclei \cite{PRL78:436:1997,PR315:231:1999,PR414:165:2005}. The study of symmetries in resonant states is certainly an interesting topic. The authors of Ref. \cite{PRL109:072501:2012} showed that the PSS in single particle resonant states in nuclei is exactly conserved, i.e., the pseudospin doublets with different quantum numbers $\kappa$ e $-\kappa+1$ have the same resonant state with energy $E_{\mathrm{res}}$ and width $\Gamma_{\mathrm{res}}$. That novel result were illustrated for single particle resonances in spherical square-well and Woods--Saxon potentials. Additionally,
the Dirac equation exhibit spin symmetry (SS) when the vector and
scalar potentials have the same magnitude and was used for explain the small
spin-orbit splitting in hadrons \cite{PRL86:204:2001}. Recently, a comprehensive review of the progress on the PSS and SS in various systems and potentials have been reported in \cite{PR570:1:2015}, including extensions of the PSS study from stable to exotic nuclei, from non-confining to confining potentials, from local to non-local potentials, from central to tensor potentials, from bound to resonant states, from nucleon to anti-nucleon spectra, from nucleon to hyperon spectra, and from spherical to deformed nuclei.

The four-dimensional Dirac equation with a mixture of spherically symmetric
scalar, vector and tensor interactions can be reduced to the two-dimensional
Dirac equation with a mixture of scalar, vector and pseudoscalar couplings
when the fermion is limited to move in just one direction ($p_{y}=p_{z}=0$) 
\cite{STRANGE1998}. In this restricted motion the scalar and vector
interactions preserve their Lorentz structures, while the tensor interaction
becomes a pseudoscalar. This kind of dimensional reduction is very useful
because the two-dimensional version of the Dirac equation can be thought as
that one describing a fermion embedded in a four-dimensional space-time with
either spin up or spin down \cite{GREINER1990}. Furthermore, the absence of
angular momentum and spin-orbit interaction as well as the use of $2\times2$
matrices, instead of $4\times4$ matrices, allow us to explore the physical
consequences of the negative-energy states in a mathematically simpler and
more physically transparent way. Therefore, we can take advantage of the
simplicity of the lowest dimensionality of the space-time.

Considering mixed scalar-vector potentials the Dirac equation in (1+1)
dimensions have been investigated for a sign potential \cite{AP340:1:2014}
and smooth step potential \cite{AP346:164:2014}. In the context of mixed
scalar-vector-pseudoscalar potentials (the most general Lorentz structure),
the bound-state solutions for fermions and antifermions in (1+1) dimensions
have been studied for harmonic oscillator potential \cite{PRC73:054309:2006}%
, P\"{o}schl-Teller potential \cite{IJMPE16:3002:2007}, Cornell potential 
\cite{AP334:316:2013,AP338:278:2013} and Coulomb potential \cite%
{AP356:83:2015}. In those works, the relation between spin and pseudospin
symmetries using charge-conjugation and chiral transformation was
illustrated.

The square potentials, wells and barriers, are models widely used in
low-dimensional systems such as the quantum dots \cite{PE67:128:2015}, Dirac fermions
in graphene \cite{PE58:30:2014}, electrons in semiconductor heterostructures \cite{PRL33:827:1974} and theoretical studies \cite{PE59:192:2014,AP361:423:2015}. Besides these applications, the
well and barrier potentials are extremely examples used as toy-models in
textbooks (for example, see \cite{GREINER1990} ), further increasing its
importance in quantum mechanics.

The main motivation of this paper is the approach of the Dirac equation
(1+1) dimensional in the framework of mixed vector-scalar-pseudoscalar
square potentials. The scattering solutions furnish an oscillatory
transmission coefficient, which does not present a total reflection. The
presence of bound-state solutions are conditioned by the intensity of the
pseudoscalar potential, which possess a critical value, compared to the
mixed vector-scalar potential. Those interesting results are obtained as
solutions of an effective Schr\"{o}dinger equation for a square potential
with repulsive and attractive delta-functions situated at the borders. The
results can give us support for study PSS and SS in high
dimensionality of the space-time.

\section{The Dirac equation in (1+1) D}

The time-independent Dirac equation for a fermion of rest mass $m$ in the
background of vector ($V_{t}$), scalar ($V_{s}$) and pseudoscalar ($V_{p}$)
potentials can written as (with units in which $\hbar =c=1$ )%
\begin{equation}
H\psi =E\psi ,\quad H=\sigma _{1}p+\sigma _{3}m+\frac{%
1+\sigma _{3}}{2}\Sigma +\frac{1-\sigma _{3}}{2}\Delta +\sigma _{2}V_{p}
\end{equation}%
\noindent where $\sigma _{1},\sigma _{2}$ and $\sigma _{3}$ are the Pauli matrices and 
$\Sigma =V_{t}+V_{s},$ $\Delta =V_{t}-V_{s}.$

The Dirac equation is covariant under $x\rightarrow -x$ if \ $V_{p}$ changes
sign whereas $V_{s}$ and $V_{t}$ remain the same. This is because the parity
operator $P=\exp \left( i\varepsilon \right) P_{0}\sigma _{3}$, where $%
\varepsilon $ is a constant phase and $P_{0}$ changes $x$ into $-x$, changes
the sign of $\sigma _{1}$ and $\sigma _{2}$ but not of $\sigma _{3}$.

The charge-conjugation operation is accomplished by the transformation $\psi
_{c}=\sigma _{1}\psi ^{\ast }$ and the Dirac equation becomes $H_{c}\psi
_{c}=-E\psi _{c}$, with 
\begin{equation}
H_{c}=\sigma _{1}p+\sigma _{3}m-\frac{I+\sigma _{3}}{2}\,\Delta -\frac{%
I-\sigma _{3}}{2}\,\Sigma +\sigma _{2}V_{p}\,.  \label{eq5a}
\end{equation}

\noindent One see that the charge-conjugation operation changes the sign of
the energy and of the potentials $V_{t}$ and $V_{p}$. In turn, this means
that $\Sigma$ turns into $-\Delta$ and $\Delta$ into $-\Sigma$. Therefore,
to be invariant under charge conjugation, the Hamiltonian must contain only
a scalar potential.

The chiral operator for a Dirac spinor is the matrix $\gamma^{5}=\sigma _{1} 
$. Under the \textit{discrete chiral transformation} the spinor is
transformed as $\psi_{\chi }=\gamma ^{5}\psi $ and the transformed
Hamiltonian $H_{\chi }=\gamma ^{5}H\gamma ^{5}$ is 
\begin{equation}
H_{\chi }=\sigma _{1}p-\sigma _{3}m+\frac{I+\sigma _{3}}{2}\,\Delta +%
\frac{I-\sigma _{3}}{2}\,\Sigma +\sigma _{2}V_{p}.  \label{eq8}
\end{equation}

\noindent This means that the chiral transformation changes the sign of the
mass and of the scalar and pseudoscalar potentials, thus turning $\Sigma$
into $\Delta$ and vice versa. A chiral invariant Hamiltonian needs to have
zero mass and $V_{s}$ and $V_{p}$ zero everywhere.

The equation (1) decomposes into two first-order equations for the upper, $%
\psi _{+},$ and the lower, $\psi _{-},$ components of the spinor:%
\begin{equation}
-i\frac{d\psi _{-}}{dx}+m\psi _{+}+\Sigma (x)\psi _{+}-iV_{p}\psi _{-}=E\psi
_{+}  \label{rel1}
\end{equation}%
\begin{equation}
-i\frac{d\psi _{+}}{dx}-m\psi _{-}+\Delta (x)\psi _{-}+iV_{p}\psi _{+}=E\psi
_{-}  \label{rel2}
\end{equation}

\noindent The components of four-current are given by $J^{0}=\psi^{\dagger}%
\psi$ and $J^{1}=\psi^{\dagger}\sigma_{1}\psi$. If we use the spinor $\psi$
in terms of its components the four-current is expressed by $J^{0}=$ $|\psi
_{+}|^{2}+|\psi _{-}|^{2}$ and $J^{1}=2\mathrm{Re}(\psi _{+}^{\ast }\psi
_{-}) $ which are conserved quantities for stationary states. Then, the
Dirac spinor is normalized as $\int_{-\infty }^{+\infty }dxJ^{0}=1$, so that 
$\psi _{+}$ and $\psi _{-}$ are square integrable functions. It is clear
from the pair of coupled first-order differential equations (\ref{rel1}) and
(\ref{rel2}) that $\psi_{+}$ and $\psi_{-}$ have opposite parities if the
Dirac equation is covariant under $x\rightarrow -x$.

For $\Delta =0$ and $E\neq-m$, using the expression for $\psi_{-}$ obtained
from (\ref{rel2}) and inserting it into eq.~(\ref{rel1}) the Dirac equation
for $\psi_{+}$ becomes%
\begin{equation}
\psi _{-}=-\frac{i}{E+m}\left( \frac{d\psi _{+}}{dx}-V_{p}\psi _{+}\right)
\end{equation}%
\begin{equation}  \label{eqsc}
-\frac{d^{2}\psi _{+}}{dx^{2}}+\left[ (E+m)\Sigma +V_{p}^{2}+\frac{dV_{p}}{dx%
}\right] \psi _{+}=\left( E^{2}-m^{2}\right) \psi _{+}
\end{equation}

\noindent Therefore, the solution of the relativistic problem is mapped into
a Sturm-Liouville problem for the upper component of the Dirac spinor. As
discussed in the ref.~\cite{IJMPE16:3002:2007}, we can take advantage of the
discrete chiral transformation ($\gamma ^{5}$) and we can obtain the
solutions for $\Sigma =0 $ from the $\Delta =0$ case. This means that the
chiral symmetry is invoked to obtain the equations obeyed by $\psi _{+}$ and 
$\psi _{-}$, for $\Sigma =0 $ and $E\neq m$. They are obtained from the
previous ones by doing $\psi _{+}\leftrightarrow \psi _{-}$, $m\rightarrow -m
$, $\Sigma \rightarrow \Delta $ and $V_{p}\rightarrow -V_{p}$\thinspace .
The solutions for $\Delta =0$ with $E=-m$ and $\Sigma =0$ with $E=m$, called
isolated solutions \cite{PLA351:379:2006,AP320:56:2005,AP338:278:2013}, are
obtained directly from the original first-order equations (\ref{rel1}) and (%
\ref{rel2}). For $\Delta =0$ and $E=-m$, we obtain 
\begin{equation}
\frac{d\psi _{+}}{dx}-V_{p}\psi _{+}=0  \label{is1}
\end{equation}%
\begin{equation}
\frac{d\psi _{-}}{dx}+V_{p}\psi _{-}=-i\left( \Sigma +2m\right) \psi _{+}
\label{is2}
\end{equation}

\noindent whose solution is 
\begin{equation}  \label{is3}
\psi_{+}=N_{+}\mathrm{e}^{+v(x)}
\end{equation}
\begin{equation}  \label{is4}
\psi_{-}=\left[ N_{-}-iN_{+}I(x) \right]\mathrm{e}^{-v(x)}
\end{equation}

\noindent where $N_{+}$ and $N_{-}$ are normalization constants, and 
\begin{equation}  \label{is5}
v(x)=\int^{x}dy V_{p}(y)
\end{equation}
\begin{equation}  \label{is6}
I(x)=\int^{x}dy \left[ \Sigma(y)+ 2m\right]\mathrm{e}^{+2v(y)}.
\end{equation}

\noindent Note that this sort of isolated solution cannot describe
scattering states and inasmuch as $\psi_{+}$ and $\psi_{-}$ are normalizable
functions, the possible isolated solution implies that $V_{p}\neq0$,
therefore the presence of a pseudoscalar potential is \textit{sine qua non}
for provide isolated solutions~\cite{AP338:278:2013}.

\section{The Square Potentials}

Lets us consider%
\begin{equation}
\Sigma (x)=C_{\Sigma }g(x)\,,  \label{rel4}
\end{equation}%
\begin{equation}
\Delta =0\,,  \label{delta1}
\end{equation}%
\begin{equation}
V_{p}(x)=C_{p}g(x)  \label{rel5}
\end{equation}

\begin{equation}
g(x)=\frac{1}{2}\left[ \mathrm{sgn}(x-a)-\mathrm{sgn}(x+a)\right]
\end{equation}

\noindent where $\mathrm{sgn}(x)=x/|x|$ ($x\neq 0$) is the sign function, $%
C_{\Sigma }$ and $C_{p}$ are constants with dimensions of energy. Due to the
chiral symmetry we can focus the discussion on the $\Sigma $ case ($\Delta
=0 $). The results for the case $\Delta \neq 0,$ $\Sigma =0$ and $V_{p},$
still given by (\ref{rel5}), can be easily obtained by just changing the
signs of $m$ and $C_{p}$ in the relevant expressions.

For the potentials (\ref{rel4}), (\ref{delta1}) and (\ref{rel5}) we have not
found a normalizable isolated solution with $E=-m$, i.e., the absence of
isolated solution is because $V_{p}(x)=0$ for $|x|>a$ and therefore $\psi
_{+}$ is a constant. For the case $E\neq -m$ the Eq.~(\ref{eqsc}) takes the
form%
\begin{equation}
-\frac{d^{2}\psi _{+}}{dx^{2}}+V_{\mathrm{eff}}(x)\psi _{+}=k^{2}\psi _{+}
\label{eqsc2}
\end{equation}

\noindent where 
\begin{equation}
V_{\mathrm{eff}}(x)=\left[ C_{\Sigma }(E+m)-C_{p}^{2}\right] g(x)+\frac{C_{p}%
}{2}\left[ \delta (x-a)-\delta (x+a)\right] ,  \label{eff}
\end{equation}%
$\delta (x)$ is the Dirac delta function and the effective energy particle
is given by $k^{2}/2m$, with $k^{2}=E^{2}-m^{2}$. We note that the effective
potential distinguishes particles and anti-particles if $C_{\Sigma }\neq 0$,
and so we can not expect a symmetric spectrum with respect to $E$.

Let us introduce the parameter $r\equiv V_{\text{eff}}(\left\vert
x\right\vert <a)$, where 
\begin{equation}
V_{\text{eff}}(\left\vert x\right\vert <a)=C_{p}^{2}-C_{\Sigma }(E+m)\,.
\label{parar}
\end{equation}

\begin{figure}[!ht]
\centering
\includegraphics[scale=0.4]{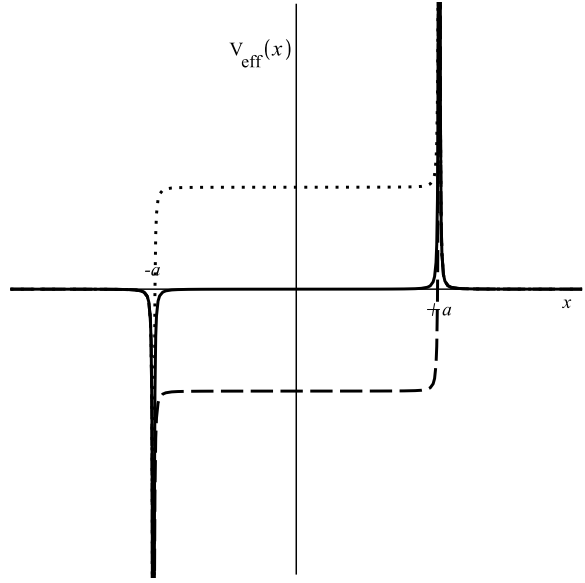}
\caption{Profiles for effective potential $V_{\mathrm{eff}}(x)$ for $r>0$
(pointed line), $r=0$ (thin line) and $r<0$ (dashed line).}
\label{perfis}
\end{figure}

\noindent The parameter $r$ characterizes three different profiles for the
effective potential as illustrated in figure (~\ref{perfis}). If $r<0$, the
effective potential consists in a finite square well potential at the region 
$-a<x<+a$ with attractive and repulsive delta function situated at $x=-a$
and $x=+a$, respectively. If $r=0$, the effective potential consists in a
double delta function potential with attractive and repulsive delta function
situated at $x=-a$ and $x=+a$, respectively. Finally, if $r>0$, the
effective potential consists in a finite square barrier potential at the
region $-a<x<+a$ with attractive and repulsive delta function situated at $%
x=-a$ and $x=+a$, respectively.

\subsection{Scattering States}

We focused our attention to the scattering states solutions that describes a
fermion moving from left to right. In this way, $\psi (x\rightarrow -\infty
) $ describes an incident wave moving to the right and a reflected wave
moving to the left, and $\psi (x\rightarrow -\infty )$ describes a
transmitted wave moving to the right or an evanescent wave. The upper
components for scattering states are written as

\begin{equation}
\psi _{+}(x)=\left\{ 
\begin{tabular}{cc}
$A_{+}\exp (ikx)+A_{-}\exp (-ikx)$ & for $x<-a$ \\ 
$B_{+}\exp (i\eta x)+B_{-}\exp (-i\eta x)$ & for $\left\vert x\right\vert <a$
\\ 
$D_{\pm }\exp (ikx)$ & for $x>+a$%
\end{tabular}%
\right.  \label{psi}
\end{equation}%
where%
\begin{equation}
\eta =\sqrt{k^{2}-r}\text{ \ \ }
\end{equation}%

\noindent The group velocity of the waves described above is given by

\begin{equation}
v_{g}=\frac{dE}{dk}=\pm \frac{k}{\sqrt{k^{2}+m^{2}}}
\end{equation}%

\noindent where the double signal is related to propagation direction. For both range
the probability current densities are given by

\begin{equation}
J^{1}(x\rightarrow -\infty )=\frac{2k}{E+m}(\left\vert A_{\pm }\right\vert
^{2}-\left\vert A_{\mp }\right\vert ^{2})\text{ \ \ \ \ \ \ for \ \ \ \ \ \ }%
E\gtrless -m  \label{cu1}
\end{equation}%

\noindent and%
\begin{equation}
J^{1}(x\rightarrow +\infty )=\pm \frac{2k}{E+m}\left\vert D_{\pm
}\right\vert ^{2}\text{ \ \ \ for \ \ \ \ \ \ }E\gtrless -m  \label{cu2}
\end{equation}%

\noindent Note that $J^{1}(-\infty )=J_{inc}-J_{ref}$ and $J^{1}(+\infty )=J_{trans}$,
where $J_{inc}$, $J_{ref}$ and $J_{trans}$ are nonnegative quantities
characterizing the incident, reflected and transmitted waves, respectively.
If $E+m>0$, then $A_{+}e^{+ikx}$ ($A_{-}e^{-ikx}$) will describe the
incident (reflected) wave, and $D_{-}=0$. On the other hand, if $E+m<0$,
then $A_{-}e^{-ikx}$ ($A_{+}e^{+ikx}$) will describe the incident
(reflected) wave, and $D_{+}=0$.

To determinate the reflection and transmission coefficients, we use the
probability current densities given by (\ref{cu1}) and (\ref{cu2}). The $x$%
-independent probability current allow us to define the reflection and
transmission coefficients as 
\begin{equation}
R_{\pm }=\left\vert r_{\pm }\right\vert ^{2}\text{ \ ,\ \ }T_{\pm
}=\left\vert t_{\pm }\right\vert ^{2}\text{ \ \ \ with \ \ \ \ }R_{\pm
}+T_{\pm }=1
\end{equation}%

\noindent where the quantities 
\begin{equation}
r_{\pm }=\frac{A_{\mp }}{A_{\pm }}\text{ \ \ \ \ \ \ and \ \ \ }t_{\pm }=%
\frac{D_{\pm }}{A_{\pm }}
\end{equation}%
\noindent are called reflection and transmission amplitudes, respectively.

We demand that $\psi _{+}$ be continuous at $x=\pm a$, that is 
\begin{equation}
\lim_{\sigma \rightarrow 0}\psi _{+}(x)\mid _{x=\pm a-\sigma }^{x=\pm
a+\sigma }=0\,.  \label{cont}
\end{equation}

\noindent Moreover, the effect due to the delta function potential on $d\psi
_{+}/dx$ in the neighborhood of $x=\pm a$ can be evaluated by integrating (%
\ref{eqsc2}) from $\pm a-\sigma $ to $\pm a+\sigma $ and taking the limit $%
\sigma \rightarrow 0$. Thereby, we obtain 
\begin{equation}
\lim_{\sigma \rightarrow 0}\frac{d\psi _{+}(x)}{dx}\mid _{x=\pm a-\sigma
}^{x=\pm a+\sigma }=\pm \frac{C_{p}}{2}\psi _{+}(x=\pm a) \,. \label{discont}
\end{equation}%
\noindent With $\psi _{+}(x)$ given by (\ref{psi}), conditions (\ref{cont}) and (\ref%
{discont}) imply that

\begin{equation}
B_{+}e^{-i\eta a}+B_{-}e^{i\eta a}=A_{+}e^{-ika}+A_{-}e^{ika}
\end{equation}
\begin{equation}
D_{\pm }e^{ika}=B_{+}e^{i\eta a}+B_{-}e^{-i\eta a} 
\end{equation}
\begin{equation}
ikD_{\pm }e^{ika}-i\eta \left(B_{+}e^{i\eta a}-B_{-}e^{-i\eta a}\right)=%
\frac{C_{p}}{2}\left(B_{+}e^{i\eta a}+B_{-}e^{-i\eta a}\right) 
\end{equation}
\begin{equation}
\begin{split}
i\eta \left(B_{+}e^{-i\eta a}-B_{-}e^{i\eta
a}\right)-& ik\left(A_{+}e^{-ika}-A_{-}e^{ika}\right)=\\
&-\frac{C_{p}}{2}%
\left(B_{+}e^{-i\eta a}+B_{-}e^{i\eta a}\right) 
\end{split}
\end{equation}

\noindent Omiting the algebraic details, we obtain the relative amplitudes
\begin{equation}
\frac{B_{+}}{A_{\pm }}=\frac{i\eta k\left( \eta +k+iC_{p}/2\right)
e^{-i(k+\eta )a}}{2d} 
\end{equation}
\begin{equation}
\frac{B_{-}}{A_{\pm }}=\frac{k^{2}\eta \left( \eta -k-iC_{p}/2\right)
e^{i(\eta -k)a}}{2d}
\end{equation}
\begin{equation}
\frac{A_{\mp }}{A_{\pm }}=\frac{i\eta k\left[ \eta ^{2}-\left(
k+iC_{p}/2\right) ^{2}\right] \sin (4\eta a)e^{-2ika}}{2d} 
\end{equation}
\begin{equation}\label{pii}
\frac{D_{\pm }}{A_{\pm }}=\frac{e^{-2ika}}{d} 
\end{equation}

\noindent where%
\begin{equation}
d\equiv \cos (2\eta a)-\sin (2\eta a)f(k)
\end{equation}%

\noindent and%
\begin{equation}
f(k)\equiv i\frac{k^{2}+C_{p}^{2}/4+\eta ^{2}}{2\eta k}\,.
\end{equation}%

\noindent The equation (\ref{pii}) given us the transmission coefficient
\begin{equation}
T=\frac{1}{1+\sin ^{2}(2\eta a)\left[ \left\vert f(k)\right\vert ^{2}-1%
\right] -2\sin (4\eta a)\mathrm{\mathrm{Re}}\left[ f(k)\right] }\,.
\end{equation}

The scattering process is only possible with localized energies in the range 
$\left\vert E\right\vert >m$ and then $k$ $\in $ $%
\mathbb{R}
.$ We see that both the effective potential as the transmission coefficient
depends on the energy and then the transmission coefficient is not symmetric
with respect to $E$. We can see that the transmission increases with energy,
namely, $\left\vert E\right\vert \rightarrow \infty $ $T\rightarrow 1.$ The
transmission coefficient provides us the necessary condition to obtain the
resonant states

\begin{equation}
\eta =\frac{\left( N+1\right) \pi }{2a},\text{ \ \ \ \ where \ }N=0,1,2,3..
\end{equation}%

\noindent Therefore, the energies of resonant states can be writen as%
\begin{equation}
E=\pm \sqrt{\frac{\left( N+1\right) ^{2}\pi ^{2}}{4a^{2}}+\left( m-C_{\Sigma
}\right) ^{2}+\frac{C_{p}^{2}}{4}}-C_{\Sigma }
\end{equation}%

\noindent and in the limit $N\rightarrow \infty $, we obtain
\begin{equation}
E=\pm \frac{N\pi }{2a}\,.
\end{equation}

\begin{figure}[tbh]
\centering
\includegraphics[scale=0.4]{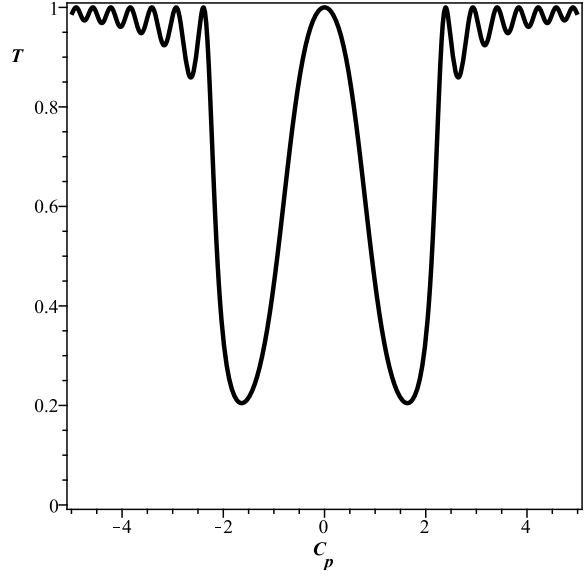}
\caption{Transmission coefficient for $r=0$, $C_{\Sigma }=2$, $m=a=1$.}
\label{transm0}
\end{figure}

For this case $r=0$ the energy is fixed ($E=C_{p}^{2}/C_{\Sigma }-m$) and we
can see that the pseudoscalar potential influence to transmission
coefficient as showed in figure \ref{transm0}. From figure \ref{transm0}, we
note an oscillatory behavior and no full reflection, as expected. As $T$
does not depend on the sing of $C_{p}$ the transmission coefficient have the
same values from both barrier and well pseudoscalar potentials.

\begin{figure}[!htb]
\centering
\includegraphics[scale=0.4]{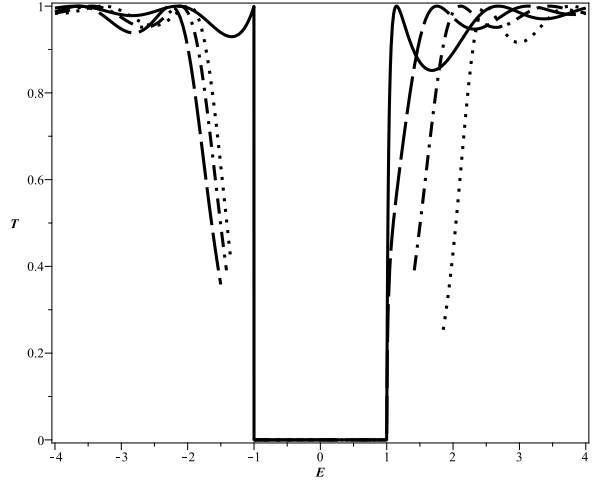}
\caption{Transmission coefficient for $m=a=1$; $C_{p}=0$, $C_{\Sigma}=1$
(solid line); $C_{p}=1$, $C_{\Sigma}=0$ (solid-dot line); $C_{p}=1$, $%
C_{\Sigma}=0.5$ (long dashed line); $C_{p}=1$, $C_{\Sigma}=-0.5$ (pointed
line).}
\label{transm}
\end{figure}

Some profiles for transmission coefficient are show in the figure \ref%
{transm} for differents values of $C_{p}$ and $C_{\Sigma}$. We can see that
there is not full reflection for all cases. Further, from figure \ref{transm}
we can note that curves of the transmission coefficient for negative energy
are closer between them in comparasion to curves for positive energy. As
expected, for very high energy, $T\rightarrow 1$ and also observe that the
pseudoscalar potential maintains the oscillatory behavior of $T$.

Also, we can observe that the energies from resonant states is much similar
to bound state energies for an infinity double-step potential

\begin{equation}
V_{p,\Sigma }(x)=\left\{
\begin{tabular}{ccc}
$-\infty $ &  & $x<-a$ \\
$C_{p,\Sigma }$ &  & $x\leq |a|$ \\
$+\infty $ &  & $x>a$%
\end{tabular}%
\right.
\end{equation}

\noindent in the limit $N\rightarrow \infty $
\begin{equation}
E=\pm \sqrt{\frac{\left( N+1\right) ^{2}\pi ^{2}}{4a^{2}}+\left( m+\frac{%
C_{\Sigma }}{2}\right) ^{2}+C_{p}^{2}}+\frac{C_{\Sigma }}{2}.
\end{equation}

The similarity is because the bound state energies corresponds approximately
to real part of the resonant energies obtained from the poles of the
transmission amplitude for a square well \cite{BAYM1969}. A study on
correspondence between behavior of $T$ and the bound-state energies for
nonrelativistic square wells and barriers was done by Maheswari and
collaborators \cite{AJP78:412:2010}. But the authors make some mistakes,
corrected by Ahmed \cite{AJP79:682:2011}, which also discusses some criteria to find
oscillatory $T$ for a potential class.

\subsection{Bound States}

The bound-states solutions ($\left\vert E\right\vert <m$) can be obtained
from (\ref{psi}) using the prescription $k=+i\left\vert k\right\vert $ and $%
A_{+}=D_{-}=0$ or $k=-i\left\vert k\right\vert $ and $A_{-}=D_{+}=0,$ i.e.,
the bound-states solutions correspond to the poles of the transmission
amplitude$.$ The conditions (\ref{cont}) and (\ref{discont}) providing the
following quantization condition

\begin{equation}\label{cq}
\frac{2\eta \left\vert k\right\vert }{\eta ^{2}+C_{p}^{2}/4-\left\vert
k\right\vert ^{2}}=\tan (2\eta a)
\end{equation}%

\noindent We note that the above condition does not depend on the sign of $C_{p}$
(barrier or square well pseudoscalar potential), and hence the energies does
not depend on the localization of the delta function. Using the
trigonometric relation $\cot (\theta )=-\tan (\theta /2)\pm \sqrt{1+\cot
^{2}(\theta )}$, the quantization condition can be rewrite as%
\begin{equation}
\tan (\eta a)=\pm \sqrt{1+\left[ f(k)\right] ^{2}}-f(k)  \label{qc}
\end{equation}%

\noindent We can see that in the absence of pseudoscalar potential ($C_{p}=0$), the
quantization condition is the same obtained for spin-$0$ bosons \cite%
{RBEF30:2306:2008}. This equivalence confirm the results obtained by P.
Alberto and collaborators \cite{PRC75:047303:2007}, which show that the spin
and pseudospin symmetries in Dirac equation produce an equivalent energy
espectra for relativistic spin-$1/2$ and spin-$0$ particles in the presence of
vector and scalar potentials.

\subsubsection{Case $r\geq 0$.}

For $r=0$ we have $\eta =\pm i\left\vert k\right\vert $, therefore (\ref{cq}) can
be writen as
\begin{equation}
\frac{2\left\vert k\right\vert ^{2}}{C_{p}^{2}/4-2\left\vert k\right\vert
^{2}}=\tanh (2\left\vert k\right\vert a).
\end{equation}

The above condition has just one solution in $\left\vert k\right\vert =0,$
therefore we do not have bound-states solutions for $r=0$. The case $r>0$
has a repulsive barrier between the two delta functions, and does not
contain bound-states solutions too. The equation (\ref{parar}) and the
condition $\left\vert E\right\vert <m$ allows us to conclude that only have
bound states for $C_{\Sigma }>0$.

\subsubsection{Case $r<0$.}

\begin{figure}[!htb]
\centering
\includegraphics[scale=0.4]{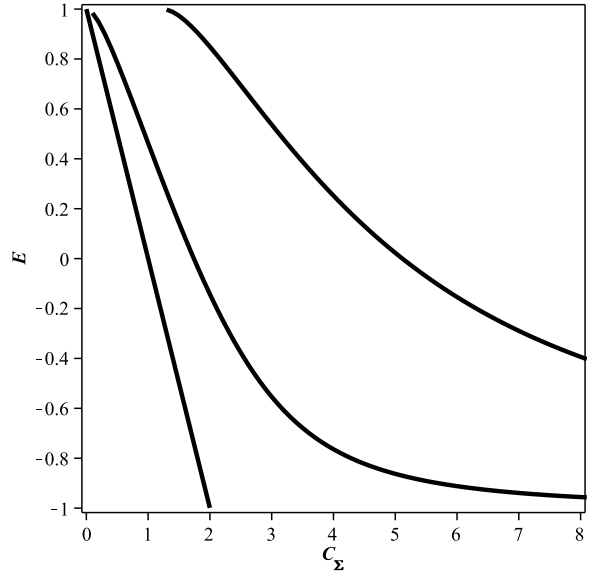}
\caption{Bound-states espectrum for $C_{p}=0$, $m=a=1$.}
\label{fig02}
\end{figure}

\begin{figure}[!htb]
\centering
\includegraphics[scale=0.4]{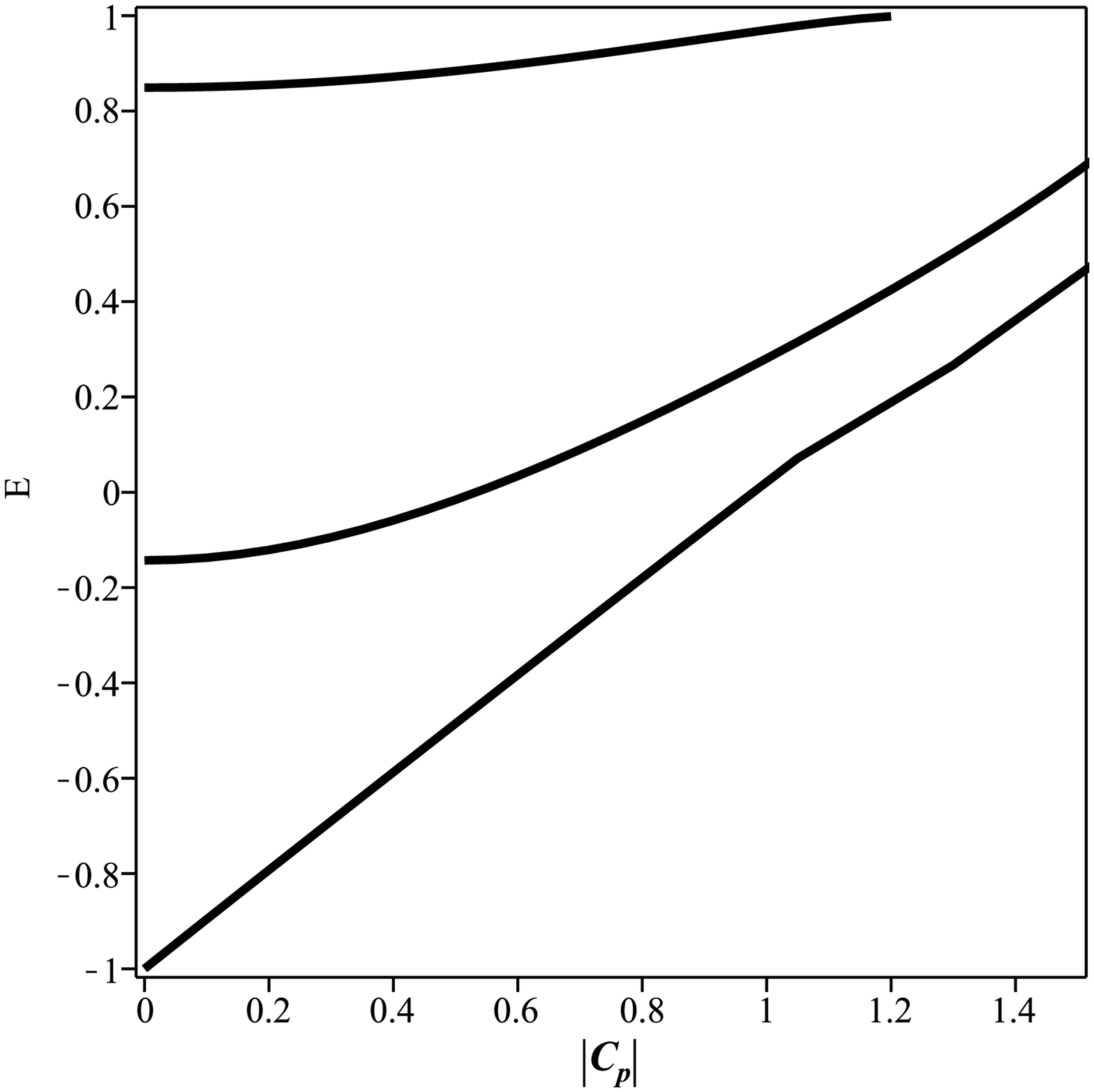}
\caption{Bound-states espectrum for $C_{\Sigma }=2$, $m=a=1$.}
\label{fig03}
\end{figure}

For $r<0$ we have the condition $0\leq C_{p}^{2}<2mC_{\Sigma }$ and the
quantization conditions provides the figures \ref{fig02} and \ref{fig03}.

The effective potential given by equation (\ref{eff}) has not defined
parity, and therefore do not know the parity of the solutions. Obviously,
the case $C_{p}=0$ show energy levels with parity defined, even (odd)
solutions to negative (positive) sign in (\ref{qc}), the same behavior is
found in reference \cite{RBEF30:2306:2008}. The ground state energy ($E_{g.s}
$), in the figure \ref{fig02} (\ref{fig03}), is given by $E_{g.s}\simeq
m-C_{\Sigma }$ ($E_{g.s}\simeq m-C_{p}\mathrm{sgn}(C_{p})$). Therefore we
obtain the constrain $C_{\Sigma }+C_{p}\mathrm{sgn}(C_{p})=2m$ for the
low-energy state. As we already know, there are always bound-states
solutions (at least one) if the intensity of the pseudoscalar barrier or
well not exceeds the critical value $C_{p}^{critic}=\sqrt{2mC_{\Sigma }}$,
for $C_{\Sigma }\neq 0$.

\section{Conclusions}

The states of fermions in the framework of mixed vector-scalar-pseudoscalar
square potentials was investigated. The condition for spin 
symmetry ($\Delta=0$) enables to
decouple the Dirac equation in an effective Schr\"{o}dinger equation with a square potential
with repulsive and attractive delta-functions situated at the borders for the upper component; and the lower component was expressed in term of the upper component in a simple way. An oscillatory transmission
coefficient and resonant states energy were obtained. We showed that
existence of bound-state solutions are conditioned by the intensity of the
pseudoscalar potential, which posses a critical value $C_{p}^{critic}=\sqrt{%
2mC_{\Sigma }}$ for $C_{\Sigma }\neq 0$. In the absence of pseudoscalar
potential, we obtain the same spectrum for spinless particles \cite{RBEF30:2306:2008}, confirming
the predictions of Ref. \cite{PRC75:047303:2007}.

This work can illustrate some general conclusions drawn in previous works about spin and pseudospin symmetries, we can obtain the solutions for $\Sigma=0$ from the $\Delta=0$ case, using the chiral transformation (changing the signs of $m$ and $C_{p}$ in the relevant expressions). Finally, it is well known that square potentials, wells and barriers are of certain interest in solid state physics, therefore our results could be applied to refine one-dimensional potential models caused by ions in a periodic crystal lattice, as the Kronig--Penney model \cite{Kronig1931}. Other possible application of our results could be in the neutron scattering on nucleus, where bound-state informations are extract for many isotopes in well \cite{PPNL8:542:2011}.

\section*{Acknowledgments}

This work was supported in part by means of funds provided by CAPES, Brazil and CNPq, Brazil, Grants No. 455719/2014--4 (Universal) and No. 304105/2014--7 (PQ). We
thank the professor A. S. de Castro, UNESP - Campus de Guaratinguet\'{a},
for valueble discussions and suggestions.

\bibliography{mybibfile_stars2}

\end{document}